\documentclass[prd,aps,twocolumn,nofootinbib,preprintnumbers]{revtex4}
\usepackage{graphicx,amsmath}
\usepackage{comment}
\usepackage{color}

\begin{document}

\newcommand{\mtrem}[1]{{\color{red} \bf $[[$ MT: #1 $]]$}}
\newcommand{\Jrem}[1]{{\color{blue} \bf $[[$ RJ: #1 $]]$}}
\newcommand{\knrem}[1]{{\color{red} \bf $[[$ KN: #1 $]]$}}

\def\beq{\begin{eqnarray}}
\def\eeq{\end{eqnarray}}
\newcommand{\gsim}{ \mathop{}_{\textstyle \sim}^{\textstyle >} }
\newcommand{\lsim}{ \mathop{}_{\textstyle \sim}^{\textstyle <} }
\newcommand{\vev}[1]{ \left\langle {#1} \right\rangle }
\newcommand{\bra}[1]{ \langle {#1} | }
\newcommand{\ket}[1]{ | {#1} \rangle }
\newcommand{\EV}{ {\rm eV} }
\newcommand{\KEV}{ {\rm keV} }
\newcommand{\MEV}{ {\rm MeV} }
\newcommand{\GEV}{ {\rm GeV} }
\newcommand{\TEV}{ {\rm TeV} }
\newcommand{\bea}{\begin{eqnarray}}   
\newcommand{\eea}{\end{eqnarray}}
\newcommand{\bear}{\begin{array}}  
\newcommand {\eear}{\end{array}}
\newcommand{\bef}{\begin{figure}}  
\newcommand {\eef}{\end{figure}}
\newcommand{\bec}{\begin{center}}  
\newcommand {\eec}{\end{center}}
\newcommand{\non}{\nonumber}  
\newcommand {\eqn}[1]{\beq {#1}\eeq}
\newcommand{\la}{\left\langle}  
\newcommand{\ra}{\right\rangle}
\newcommand{\ds}{\displaystyle}
\def\SEC#1{Sec.~\ref{#1}}
\def\FIG#1{Fig.~\ref{#1}}
\def\EQ#1{Eq.~(\ref{#1})}
\def\EQS#1{Eqs.~(\ref{#1})}
\def\GEV#1{10^{#1}{\rm\,GeV}}
\def\MEV#1{10^{#1}{\rm\,MeV}}
\def\KEV#1{10^{#1}{\rm\,keV}}
\def\lrf#1#2{ \left(\frac{#1}{#2}\right)}
\def\lrfp#1#2#3{ \left(\frac{#1}{#2} \right)^{#3}}


\preprint{UT-15-39}
\title{Gravitational waves from the first order phase transition of the Higgs field \\ 
at high energy scales}
\renewcommand{\thefootnote}{\alph{footnote}}

\author{
Ryusuke Jinno$^{a}$,
Kazunori Nakayama$^{a,b}$
and
Masahiro Takimoto$^{a}$}

\affiliation{
 $^a$Department of Physics, University of Tokyo, Tokyo 113-0033, Japan\\
 $^b$Kavli Institute for the Physics and Mathematics of the Universe, UTIAS, University of Tokyo, Kashiwa 277-8583, Japan
  }

\begin{abstract}

In a wide class of new physics models, there exist 
scalar fields that obtain vacuum expectation values of high energy scales. 
We study the possibility that the standard model Higgs field has experienced first order phase transition 
at the high energy scale due to the couplings with these scalar fields.
We estimate the amount of gravitational waves produced by the phase transition, 
and discuss observational consequences.

\end{abstract}

\maketitle

\section{Introduction}

Detection of gravitational waves (GWs) is one of the most promising tools to probe the early Universe. 
Possible cosmological sources for GWs include 
inflationary quantum fluctuations~\cite{Starobinsky:1979ty}, cosmic strings~\cite{Vilenkin:2000jqa}, 
and phase transitions~\cite{Witten:1984rs,Hogan:1986qda}.
In particular, if a first order phase transition occurs in the early Universe, 
the dynamics of bubble collision~\cite{Turner:1990rc,Kosowsky:1991ua,Kosowsky:1992rz,Turner:1992tz,Kosowsky:1992vn} 
and subsequent turbulence of the plasma~\cite{Kamionkowski:1993fg} are expected to generate GWs. 
These might be within a sensitivity of future space interferometer experiments such as 
eLISA~\cite{Seoane:2013qna}, Big Bang Observer (BBO)~\cite{Harry:2006fi} 
and DECi-hertz Interferometer Observatory (DECIGO)~\cite{Seto:2001qf}
or even ground-based detectors such as Advanced LIGO~\cite{Harry:2010zz}, KAGRA~\cite{Somiya:2011np} and VIRGO~\cite{TheVirgo:2014hva}.

In this paper we focus on GWs from the first order phase transition associated with the spontaneous symmetry breaking of the
standard model Higgs boson.
The properties of the phase transition of the Higgs field have long been studied in the literature 
both perturbatively~\cite{
Dine:1992wr,Farakos:1994kx,Arnold:1992rz,Fodor:1994bs} 
and nonperturbatively~\cite{
Kajantie:1995kf,Karsch:1996aw,Kajantie:1996mn,Kajantie:1996qd,Gurtler:1997hr,Rummukainen:1998as,Csikor:1998eu,Aoki:1999fi}
and it was found that the first order phase transition within the standard model
does not occur unless the Higgs boson mass is smaller than $\sim 80 $ GeV.

However, new physics beyond the standard model may greatly change the situation.
For example, in singlet extensions of the standard model, 
the new singlet scalar changes the Higgs potential at the origin
and it may induce strong first order phase transitions.
Actually in a wide class of new physics models, 
there exists a scalar field $\phi_{\rm NP}$ that obtains a vacuum expectation value
of the new physics scale $v_{\rm NP}$. 
One of the well-known examples is the Peccei-Quinn scalar field~\cite{Peccei:1977hh},
which solves the strong charge parity problem elegantly and
obtains a vacuum expectation value $v_{\rm NP}\sim 10^{10}$\,GeV~\cite{Kim:1986ax}.
In general, if there exists a scalar field $\phi_{\rm NP}$,
the quartic coupling term $\lambda^2|\phi_{\rm NP}|^2|H|^2$, where $H$ is the standard model Higgs field, exists.
It is natural for the coupling $\lambda$ to be nonzero
since any symmetry does not forbid this quartic coupling term.

In this paper, we take into account this quartic coupling between the scalar field $\phi_{\rm NP}$
and the standard model Higgs field $H$, and study the cosmological consequences, especially GW production. 
When the temperature of the Universe is higher than the scale of the new physics, 
both $\phi_{\rm NP}$ and $H$ are supposed to be trapped at the origins of their potential. 
As the temperature drops down to the scale of the new physics, 
the first order phase transition of the Higgs field may occur  
because the scale of the Higgs potential becomes the new physics one. 
We consider the standard model-like Higgs sector and some singlet extended models, 
and estimate the strength of GWs generated by this transition. 
Our setup is rather general and can be applied to many classes of new physics models.

In Sec.~\ref{setup}, we introduce our setup and briefly sum up the effective potential.
Then we show the thermal history of our scenario.
The properties of GWs generated by first order phase transitions are also
summarized.
In Sec.~\ref{Res}, we estimate the GWs generated by the phase transition of the Higgs field.
We first consider a setup where the Higgs sector is just the standard model one.
Even in such a case, a first order phase transition occurs due to the
smallness of the quartic self coupling of the Higgs field at high temperature,
though the amplitude of produced GWs is found to be below the observational sensitivities.
Then we consider singlet extensions as an example of nontrivial Higgs sector.
In this case we find that GWs with a detectable amplitude are produced during the phase transition.
 Sec.~\ref{Dis} is devoted to conclusions.

\section{Setup}
\label{setup}

\subsection{Model}

We consider the following scalar potential
\begin{align}
	V_0&=\lambda^2(|\phi_{\rm NP}|^2-v_{\rm NP}^2-\delta_{\rm EW}^2)|H|^2+\frac{\lambda_{H}}{2}|H|^4\nonumber \\
	&\;\;\;\;+\lambda^2_{\phi}(|\phi_{\rm NP}|^2-v_{\rm NP}^2)^2+V_S\nonumber \\
	V_S&=\sum_i\frac{\lambda_{SH}^2}{2}S_i^2|H|^2+\sum_i\frac{\lambda_{S\phi}^2}{2}S_i^2|\phi_{\rm NP}|^2,
	\label{V0}
\end{align}
where $\phi_{\rm NP}$ is a new scalar field that obtains the vacuum expectation value $v_{\rm NP}$ at zero temperature,
$H$ is the standard model Higgs field and $\delta_{\rm EW}$ denotes the electroweak scale that is needed to
realize a correct electroweak scale at zero temperature. 
Note that the coefficient of $|H|^2$ must be fine-tuned if $v_{\rm NP} \gg \delta_{\rm EW}$, 
but we do not pursue the origin of this tuning.
The last term $V_S$ denotes coupling between the standard model Higgs boson or $\phi_{\rm NP}$ and additional singlet real scalars $S_i(i=1,2,..,N_S)$. 
We take this form of $V_S$ as an example of a nontrivial Higgs sector.
We neglect the mass terms for $S_i$ because heavy fields do not affect the effective potential of the Higgs field.\footnote{
If the mass of $S_i$ is of the order of the temperature at the phase transition, the strength of produced GWs
are affected by some factor. We do not consider such effects for simplicity.
}
We assume universal couplings $\lambda_{SH}$ and $\lambda_{S\phi}$ for simplicity.

In this setup, the first order phase transition of the Higgs field may occur as follows.
When the temperature of the Universe is much higher than the new physics scale,
$\phi_{\rm NP}$ is trapped at the origin due to the thermal effects.
The effective potential for the Higgs field can be approximately written as
\begin{align}
	V_{\rm eff}(T,H)\simeq -\lambda^2v ^2_{\rm NP}|H|^2+\frac{\lambda_{H}(T)}{2}|H|^4+V_{\rm th}(T,H),
\end{align}
where $V_{\rm th}(T,H)$ denotes the thermal potential and $\lambda_{H}(T)$ indicates the coupling value at the temperature $T$.
Note that, in this case,
the Higgs field has a negative mass term that will trigger the phase transition of
the Higgs field at a temperature around the new physics scale.
In some parameter spaces, the transition becomes a first order one and the GWs are generated.
In Sec.~\ref{Res}, we consider the phase transition of the Higgs field in detail.

In the rest of this section, we explain the potential including thermal corrections and the thermal history of this model. 
Then we sum up the properties of GWs generated by the first order phase transition.

\subsection{Higgs potential}

When $\phi_{\rm NP}$ is trapped at the origin, the effective potential for the Higgs field at the one-loop level 
at $T$ can be written as
\begin{align}
	V
	&=-\lambda^2v ^2_{\rm NP}|H|^2+\frac{\lambda_{H}(T)}{2}|H|^4+V_{\rm CW}(H)+V_{\rm th}(T,H) \nonumber \\
	&\;\;\;\;+c(T),
	\label{eq_V}
\end{align}
where $V_{\rm CW}$ denotes the Coleman-Weinberg potential,
\begin{align}
	V_{\rm CW}(H)
	&=
	\frac{6}{64\pi^2}m_W(H)^4\left[\log(m_W(H)^2/\mu^2)-\frac{5}{6} \right]\nonumber \\
	&\;\;\;\;+\frac{3}{64\pi^2}m_Z(H)^4\left[\log(m_Z(H)^2/\mu^2)-\frac{5}{6} \right]\nonumber \\
	&\;\;\;\;-\frac{12}{64\pi^2}m_t(H)^4\left[\log(m_t(H)^2/\mu^2)-\frac{3}{2} \right]\nonumber \\
	&\;\;\;\;+\frac{N_S}{64\pi^2}m_S(H)^4\left[\log(m_S(H)^2/\mu^2)-\frac{3}{2} \right],\\
	m_W(H)&=\frac{g_2}{\sqrt{2}}|H|,~m_Z=\sqrt{\frac{g_2^2+g'^2}{2}}|H|,\nonumber \\
	m_t(H)&=y_t|H|,~m_S(H)=\lambda_{SH}|H|.
\end{align}
with $g_2,~g',~y_t,$ being the weak gauge coupling, 
the hypercharge gauge coupling and top Yukawa coupling, respectively. 
Also, $N_S$ denotes the number of the real singlets. 
We omitted the contributions from $\phi_{\rm NP}$ and the Higgs boson.
Throughout this paper, we set the renormalization scale $\mu$ to be 
the temperature at the phase transition.
The thermal contribution $V_{\rm th}(T,H)$ can be written as
\begin{align}
	V_{\rm th}(T,H)
	&=
	3V_{\rm th}^B(m_W(H)/T,T)+\frac{3}{2}V_{\rm th}^B(m_Z(H)/T,T)\nonumber \\
	&\;\;\;\;+6V_{\rm th}^{F}(m_t(H)/T,T)+\frac{N_S}{2}V_{\rm th}^B(m_S(H)/T,T)\nonumber \\
	&\;\;\;\;+V_{\rm daisy},\\
	V_{\rm daisy}&=-\frac{T}{6\pi}\Bigl[(M_W(H,T)^3-m_W(H)^3)\nonumber \\
	&\;\;\;\;\;\;\;\;\;\;\;\;\;\;+\frac{1}{2}(M_Z(H,T)^3-m_Z(H)^3)\nonumber \\
	&\;\;\;\;\;\;\;\;\;\;\;\;\;\;+\frac{N_S}{2}(M_S(H,T)^3-m_S(H)^3)
	\Bigr],\\
	M_W(T,H)^2&=m_W(H)^2+\frac{11g_2^2T^2}{6},\\
	M_Z(T,H)^2&=m_Z(H)^2+\frac{11(g_2^2+g'^2)T^2}{6},\\
	M_S(T,H)^2&=m_S(H)^2+\frac{\lambda_{SH}^2T^2}{6},\\
	V_{\rm th}^{B/F}(x,T)&\equiv \pm \frac{T^4}{\pi^2}\int_0^{\infty}dz~z^2\log\left[1\mp e^{-\sqrt{z^2+x^2}}\right],
\end{align}
where $V_{\rm daisy}$ denotes so-called daisy subtraction~\cite{Arnold:1992rz}.
We neglect the effects from the coupling $\lambda_{S\phi}$ by assuming $\lambda_{S\phi} < \lambda_{SH}$ for simplicity.
The Higgs-independent term $c$ in Eq.~(\ref{eq_V}) is introduced to set the symmetric minimum of the Higgs potential 
to have $V=0$. This is just for the bounce calculation in later sections, and 
does not affect the dynamics of the model.

\subsection{Thermal history}

\begin{figure}
  \centering
  \includegraphics[width=\columnwidth]{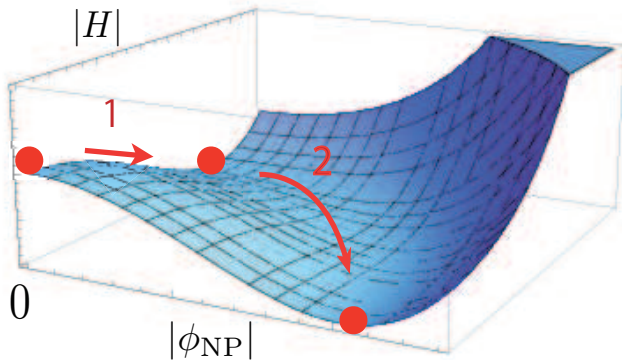}
  \caption{Schematic picture of the zero temperature potential.
  First, both $H$ and $\phi_{\rm NP}$ sit at the origin. The phase transition labeled as ``1'' in the figure occurs at $T=T_{H}^{\rm PT}$.
  Then the next phase transition, labeled as ``2'' in the figure, occurs at $T=T_{\phi_{\rm NP},H\neq 0}^{{\rm PT}}$.}
  \label{fig:pot}
\end{figure}

Let us consider the thermal history of this model.
At high temperature, both $\phi_{\rm NP}$ and $H$ obtain the thermal masses, which we parametrize as
$y_{\phi}^2T^2|\phi_{\rm NP}|^2$ and $y^2_HT^2|H|^2$
respectively. 
The parameter $y_\phi$ depends on the coupling of $\phi_{\rm NP}$ with other particles 
and therefore we treat it as a free parameter.
On the other hand,
$y_H$ is $\mathcal{O}(1)$ parameter depending on the standard model gauge couplings and Yukawa couplings.
When the temperature of the Universe is high enough, both $\phi_{\rm NP}$ and $H$ are 
trapped at the origins due to the thermal mass terms.
At the origins with the temperature $T$, the effective masses of $\phi_{\rm NP}$ and $H$ can be written as
\begin{align}
	m^2_{\phi_{\rm NP},{\rm eff}}(T)&=y_{\phi}^2T^2-2\lambda^2_{\phi}v_{\rm NP}^2,\\
	m^2_{H,{\rm eff}}(T)&=y_{H}^2T^2-\lambda^2v_{\rm NP}^2.
\end{align}
From these expressions, we get the transition temperature of $\phi_{\rm NP}$ and $H$
\begin{align}
	T_{\phi_{\rm NP}}^{\rm PT}&\simeq \lambda_\phi v_{\rm NP}/y_{\phi},\\
	T_{H}^{\rm PT}&\simeq \lambda v_{\rm NP}/y_{H}.
\end{align}
From now on we consider the case with $T_{H}^{\rm PT}>T_{\phi_{\rm NP}}^{\rm PT}$.
With this condition the phase transition of the Higgs field occurs first,
which corresponds to the arrow labeled as ``1'' in Fig.~\ref{fig:pot}. 
As the temperature drops down further, the phase transition of $\phi_{\rm NP}$ occurs.
This corresponds to the arrow labeled as ``2'' in the same figure.
After this second phase transition the Higgs field is trapped at the origin again until the temperature 
drops to the electroweak scale.
This final electroweak phase transition proceeds just in the same way as that in the standard model.

Now let us consider the entropy injection caused by the secondary phase transition of $\phi_{\rm NP}$,
which might potentially significantly dilute GWs produced by the preceding phase transition of the Higgs field.
After the phase transition of the Higgs field at $T_H^{\rm PT}$,
the Higgs field settles down to the temporal minimum $|H|^2 \simeq \lambda^2 v_{\rm NP}^2/\lambda_H$
and $\phi_{\rm NP} = 0$, denoted by the red circle between the two arrows in Fig.~\ref{fig:pot}.
The effective mass of $\phi_{\rm NP}$ at the temporal minimum
can be written as  
\begin{align}
	m_{\phi_{\rm NP},{\rm eff}}^2&=y_{\phi}^2T^2-\left(2\lambda_\phi^2-\frac{\lambda^4}{\lambda_H}\right)v_{\rm NP}^2
	\nonumber \\&\equiv y_\phi^2T^2-\epsilon^2v_{\rm NP}^2.
\end{align}
We need $\epsilon^2>0$ to ensure that the present electroweak symmetry breaking vacuum is the true vacuum.
If this condition is satisfied, $\phi_{\rm NP}$ becomes tachyonic at the temperature
\begin{align}
	T_{\phi_{\rm NP},H\neq 0}^{\rm PT}=\frac{\epsilon}{y_\phi}v_{\rm NP}.
\end{align}
The phase transition of $\phi_{\rm NP}$ happens at around this temperature
and the system relaxes to $|\phi_{\rm NP}| = v_{\rm NP}$ and $H=0$ until the temperature drops down to the electroweak scale.
We parametrize the ratio of the two phase transition temperatures as
\begin{align}
\label{cond:jun}
	\frac{T_{\phi_{\rm NP},H\neq 0}^{{\rm PT}}}{T_{H}^{{\rm PT}}}\equiv \eta=\frac{\epsilon y_H}{\lambda y_\phi }<1.
\end{align}
On the other hand, the vacuum energy density of $\phi_{\rm NP}$ field, $V_{\phi}$, after the phase transition of $H$ is given by
\begin{align}
	V_{\phi}=\epsilon^2 v_{\rm NP}^4.
\end{align}
The energy ratio between the vacuum energy $V_{\phi}$ and the radiation component $\rho_{\rm rad}$
at the time of the phase transition of $\phi_{\rm NP}$ can be written as
\begin{align}
	\Delta\equiv \frac{\rho_{\rm rad}}{V_{\phi}}=\frac{g_\ast\pi^2}{30}\frac{\eta^4\lambda^4}{y_H^4\epsilon^2},
\end{align}
where $g_\ast$ denotes the effective degrees of freedom of relativistic particles.
If the condition $\epsilon\lesssim \eta^2\lambda^2/y_H^2$ is satisfied,
$\Delta$ exceeds one and the entropy injection due to the phase transition of $\phi_{\rm NP}$ is safely neglected.\footnote{
After the phase transition, $\phi_{\rm NP}$ starts to oscillate around $\phi_{\rm NP}\sim v_{\rm NP}$.
The $\phi_{\rm NP}$ oscillation is supposed to dissipate very soon at high temperature~\cite{Moroi:2013tea}.
}

To summarize this subsection, the thermal history we consider is the following.
The phase transition of the Higgs field occurs first, when the temperature of the Universe becomes $T_H^{\rm PT}$. 
Then the phase transition of $\phi_{\rm NP}$ occurs at 
$T=T^{\rm PT}_{\phi_{\rm NP},H\neq 0}$. 
These phase transitions occur at a temperature much higher than the electroweak scale.
After these transitions, the Higgs field settles down to the origin until the temperature becomes the electroweak scale.
The final electroweak phase transition is just the same as in the standard model: it is a crossover transition, and hence 
no additional GWs are produced. Thus we consider GWs produced at $T\simeq T_H^{\rm PT}$.
This scenario is realized if $T_{H}^{\rm PT}>T_{\phi_{\rm NP}}^{\rm PT}$ and $\epsilon^2>0$ hold. 
Also, the entropy injection caused by the transition of $\phi_{\rm NP}$ can be
neglected if $\Delta\gtrsim1$. 
We assume that these three conditions are satisfied throughout this paper.

\subsection{First order phase transition and gravitational waves}

In this subsection, we briefly summarize the properties of GWs produced by a first order phase transition.

In first order phase transitions, there are two main sources for GW production: 
bubble collisions and turbulence~\cite{Kamionkowski:1993fg}.\footnote{
However, see~\cite{Hindmarsh:2013xza,Hindmarsh:2015qta} for sound waves after bubble collisions as another source. 
Here we simply consider the two sources explained in the main text.
}
After bubbles are nucleated, they expand, storing more and more energy in their walls 
in the form of gradient and kinetic energy. 
This energy is converted to GW radiation when these bubbles collide and the spherical symmetry of each bubble is broken. 
On the other hand, bubbles induce turbulent bulk motion of the fluid, and this is known as another strong source for GWs.

The frequency and amplitude of GWs from these two sources take different values depending on the 
combustion mode of the bubble walls. 
Two different types of combustion are known, detonation and deflagration.
The former occurs when the bubble front expands faster than the sound speed,
and the bubble front is followed by the rarefaction front propagating with the sound speed.
In this case a relatively large amplitude of GWs is expected from both bubble collision and turbulence,
and we assume the transition occurs via this combustion mode in the following.
On the other hand, when the speed of bubble walls is slower than the sound speed,  
the bubble front is preceded by the shock front. This is called deflagration, 
and the GW amplitude from bubble collisions is thought to be relatively suppressed in this case~\cite{Kamionkowski:1993fg}. 
However, also in this case, turbulent motion of the fluid can be a source for GWs.

The most important parameters in determining the properties of the GW spectrum are 
the ones traditionally called $\alpha$ and $\beta$.
The former is defined as the ratio of the latent heat density to the radiation energy density at the transition,
and is given by
\begin{align}
\alpha
&= \frac{\epsilon_*}{\frac{\pi^2}{30}g_*T_*^4},
\end{align}
where $T_*$ and $\epsilon_*$ are the temperature and latent heat density at the transition, respectively.
The other quantity $\beta$ is defined by the nucleation rate per unit volume 
\begin{align}
\Gamma 
&= \Gamma_0 \exp (\beta t).
\end{align}
We explain how to calculate $\alpha$ (especially $\epsilon_*$) 
and $\beta$ from the scalar potential in the next subsection.

GW spectrum from first order phase transitions can be expressed in terms of these parameters.
Both analytical and numerical calculations of the GW frequency and amplitude have been carried out in the literature~\cite{
Kosowsky:1991ua,Kosowsky:1992rz,Kosowsky:1992vn,Kamionkowski:1993fg,
Hindmarsh:2013xza,Hindmarsh:2015qta,Kalaydzhyan:2014wca,Caprini:2007xq,Caprini:2009yp,Caprini:2010xv,Nicolis:2003tg,Caprini:2006jb,Huber:2008hg,Giblin:2014qia,Dolgov:2002ra,Gogoberidze:2007an}.

\subsubsection{Bubble collision}

For GWs from bubble collisions, we refer to the expressions in~\cite{Huber:2008hg}, 
which are applicable to detonation bubbles:
\begin{align}
	&f_{\rm peak}
	\simeq 17 \; \left(\frac{f_*}{\beta}\right) \left(\frac{\beta}{H_\ast}\right)
	\left(\frac{T_\ast}{10^8~\text{GeV}}\right)\left(\frac{g_\ast}{100}\right)^{\frac{1}{6}}~[\text{Hz}],
	\label{eq_f_coll} \\
	&h_0^2\Omega_{\rm GW}(f_{\rm peak})
	\simeq 1.7 \times 10^{-5} \nonumber \\
	&\;\;\;\;\;\;\;\;\;\;\;\;\;\;\;\;\;\;\;\;\;\;\;\; \times 
	\kappa^2 \Delta \left( \frac{\beta}{H_*} \right)^{-2}
	\left(\frac{\alpha}{1+\alpha}\right)^2 
	\left( \frac{g_*}{100} \right)^{-\frac{1}{3}},
	\label{eq_h2Omega_coll}
\end{align}
where $H_*$ and $g_*$ are the Hubble parameter and the effective degrees of freedom in the thermal bath
at the phase transition, respectively, and  $h_0$ is the reduced Hubble constant at present.
Also, $\kappa$ is the efficiency factor, the fraction of the latent heat which goes into kinetic energy of the fluid~\cite{Kamionkowski:1993fg}
\begin{align}
\kappa
&= \frac{1}{1+0.715\alpha} \left[ 0.715\alpha + \frac{4}{27}\sqrt{\frac{3\alpha}{2}}\right].
\end{align} 
In addition, $\Delta$ and$f_*/\beta$ are given by
\begin{align}
\Delta
&= \frac{0.11v_b^3}{0.42 + v_b^2}, \\
\frac{f_*}{\beta}
&= \frac{0.62}{1.8 - 0.1v_b + v_b^2}.
\end{align}
Here $v_b$ is the bubble wall velocity, which has the following expression 
in the strong phase transitions~\cite{Steinhardt:1981ct}\footnote{
See Refs.~\cite{Bodeker:2009qy,Kozaczuk:2015owa} for more discussion on the bubble wall velocity. 
}
\begin{align}
v_b
&= \frac{1/\sqrt{3} + \sqrt{\alpha^2 + 2\alpha/3}}{1+\alpha}.
\label{eq_vb}
\end{align}

\subsubsection{Turbulence}

For the GW spectrum from turbulence we refer to~\cite{Kamionkowski:1993fg}
\begin{align}
	&f_{\rm peak}
	\simeq 2.6 \; v_b^{-1}v_0\left(\frac{\beta}{H_\ast}\right) 
	\left(\frac{T_\ast}{10^8~\text{GeV}}\right)
	\left(\frac{g_\ast}{100}\right)^{\frac{1}{6}}~[\text{Hz}],
	\label{eq_f_tur} \\
	&h_0^2\Omega_{\rm GW}(f_{\rm peak})
	\simeq 10^{-5} \left(\frac{\beta}{H_\ast}\right)^{-2} v_bv_0^6\left(\frac{g_\ast}{100}\right)^{-\frac{1}{3}}.
	\label{eq_h2Omega_tur}
\end{align}
Here, $v_0$ is the typical velocity on the length scale $v/\beta$, the largest scale on which the turbulence is driven.
In weak detonation limit $v_0\sim(\kappa \alpha)^{1/2}$ while in strong detonation limit $v_0\sim 1$,
and therefore we simply use $v_0 = {\rm Min}[(\kappa\alpha)^{1/2},1]$ in the following calculation.

\subsection{Bounce calculation}

Having explained the parameter dependence of the peak frequency and amplitude of the GW spectrum,
we now illustrate how to calculate $\alpha$ and $\beta$ from a given potential
and how to determine the transition time (or temperature).

When the order parameter of the phase transition is a real scalar field, the nucleation rate per unit volume $\Gamma$ 
is given by $\Gamma = \Gamma_0 {\mathrm e}^{-S}$, where $S$ is the Euclidean action~\cite{Coleman:1977py,Callan:1977pt}
\begin{align}
S
&= \int d\tau d^3x \left[ \frac{1}{2} \left( \frac{d\Phi}{d\tau} \right)^2 + \frac{1}{2}(\nabla \Phi)^2 + V(\Phi) \right].
\label{eq_bounceS}
\end{align}
Here $\tau$ is the Euclidean time and $\Phi$ denotes the scalar field driving the transition.
In finite temperature, the action must be periodic in $T^{-1}$ and the action must be modified to be 
$\Gamma = \Gamma_0 {\mathrm e}^{-S_3 / T}$~\cite{Linde:1981zj} where
\begin{align}
S_3(T)
&= \int d^3x \left[ \frac{1}{2} (\nabla \Phi)^2 + V(\Phi,T) \right].
\label{eq_bounceS3}
\end{align}
In our setup $\Phi$ corresponds to the (real) Higgs field, and $V$ in Eq.~(\ref{eq_bounceS3}) is 
the same as Eq.~(\ref{eq_V}).
In order to calculate the profile of the scalar field at the bubble nucleation, 
one must find the $O(3)$ symmetric solution of the equation of motion
\begin{align}
\frac{d^2 \Phi}{dr^2} + \frac{2}{r} \frac{d\Phi}{dr} - \frac{\partial V}{\partial \Phi}
&= 0,
\end{align}
where $r$ denotes the variable in the radial direction, with the boundary conditions
\begin{align}
\Phi(r = \infty)
&= \Phi_{\rm false}, \\
\frac{d\Phi}{dr} (r=0)
&= 0.
\end{align}
This solution corresponds to the one where $\Phi$ rolls down the inverse potential $-V$
from a point near the true vacuum to reach the symmetric false vacuum at $r=\infty$. 
Then the bounce action is calculated as
\begin{align}
S_3
&= \int 4\pi r^2 dr \; \left[ \frac{1}{2} \left( \frac{d\Phi}{dr} \right)^2 + V \right].
\end{align}
Since $\beta = \dot{\Gamma}/\Gamma$ at the transition from its definition,
one has
\begin{align}
\frac{\beta}{H_*}
&= \left. T \frac{d(S_3 / T)}{dT} \right|_{T=T_*}.
\end{align}
In determining the other parameter $\alpha$, one uses the expression for the latent heat density
\begin{align}
\epsilon_*
&= \left[ -V_{\rm min}(T) + T\frac{d}{dT}V_{\rm min}(T) \right]_{T=T_*}.
\end{align}
Here $V_{\rm min}(T)$ is the temperature-dependent true minimum of the effective potential of 
the scalar field which drives the phase transition.
Note that the true minimum of the potential must be set to zero by adding a constant at each time.

In addition, the transition temperature $T_*$ is evaluated by the condition~\cite{Quiros:1994dr}
\begin{align}
\left. \frac{S_3}{T} \right|_{T=T_*}
&= 137 + 4\log (100\,\text{GeV}/T_*).
\end{align}

\section{Estimate of gravitational waves}
\label{Res}

In this section, we estimate the strength of GWs generated by the first order phase transition of the Higgs field.
We estimate the temperature-dependent bounce action $S_3(T)$ by using the potential introduced in the previous section,
\begin{align}
	V&=-\lambda^2v ^2_{\rm NP}|H|^2+\frac{\lambda_{H}(T)}{2}|H|^4+V_{\rm CW}(H)+V_{\rm th}(T,H).
\end{align}
We evaluate $\lambda_H(T)$ by using the two-loop renormalization group equation 
with the Higgs mass $m_h=125.09$\,GeV~\cite{Agashe:2014kda,Aad:2015zhl}
and the top mass  $m_t=173.34$\,GeV~\cite{ATLAS:2014wva}.

In the following, we first consider the situation where the Higgs sector consists of 
the standard model Higgs boson and $\phi_{\rm NP}$,
i.e., there are no additional singlet fields $S_i$ in Eq.~(\ref{V0}).
In this case we see that the amplitude of produced GWs is too weak to detect.
Then we consider singlet extensions as an example of the nontrivial Higgs sector.
The singlet(s) have basically two effects on the strength of the Higgs phase transition: 
First, they change the shape of the Higgs thermal potential around the origin. 
Second, they change the running of the Higgs quartic coupling.
We investigate these possibilities in the following.

\subsection{Standard model-like Higgs sector}
\label{sm}

In this subsection, we consider the situation where the Higgs sector consists of 
the standard model Higgs boson and $\phi_{\rm NP}$,
i.e., there are no additional singlet fields $S_i$.
In this case, it depends on the Higgs quartic coupling $\lambda_H$ 
and the gauge couplings $g$ whether the phase transition is the first order one or not.
It is shown that the electroweak phase transition becomes
first order if the Higgs mass is small enough, $m_h\lesssim 80~{\rm GeV}$~\cite{Rummukainen:1998as}. 
Since the strength of the phase transition is determined by the combination $\lambda_H/g^2$,
this implies that the first order phase transition of the Higgs field will occur 
for $\lambda_H/g^2\lesssim 0.2$~\cite{Rummukainen:1998as}.
In our case, the phase transition occurs at high temperature of the new physics scale $T\sim \phi_{\rm NP}$ 
where the quartic coupling $\lambda_H$ is much smaller than the value at the electroweak scale,
and hence the condition $\lambda_H/g^2\lesssim 0.2$ can be easily realized.

\begin{figure}
  \centering
  \includegraphics[width=\columnwidth]{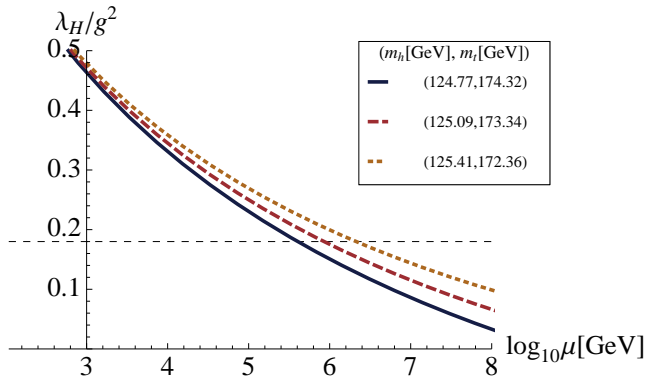}
  \caption{The temperature dependence of $\lambda_H/g^2$.
  The black-dashed line corresponds to $\lambda_H/g^2 = 0.18$.
  Each color corresponds to $(m_h,m_t) = (124.77,174.32)$ (blue), 
  $(125.09,173.34)$ (red), and $(125.41,172.36)$ (yellow).
  The left end points correspond to the transition temperature at which $\lambda_H / g^2 = 0.18$.}
  \label{fig_lHrun_SM}
\end{figure}

For concreteness, we take the criteria of the first order phase transition as
\begin{align}
	\frac{\lambda_H}{g^2}\lesssim 0.18,
\end{align}
which corresponds to the condition $m_h\lesssim70~\text{GeV}$ at the electroweak scale.
Then we estimate $\beta/H_\ast$ using the potential (\ref{eq_V}).\footnote{
Strictly speaking, near the critical point ($\lambda_H/g^2\sim 0.18$),
the strength of the GWs is supposed to be suppressed compared to our simple estimation.
For such a region, our calculation gives an upper bound of the GWs, which is already far below the observable strength. 
}
Figure~\ref{fig_lHrun_SM} shows the temperature dependence of $\lambda_H/g^2$ while varying the top quark mass.
It is seen that for $T\gtrsim 10^6$ GeV, the condition $\lambda_H/g^2\lesssim 0.18$ is satisfied and the first order phase transition will occur.

\begin{figure}
  \centering
  \includegraphics[width=\columnwidth]{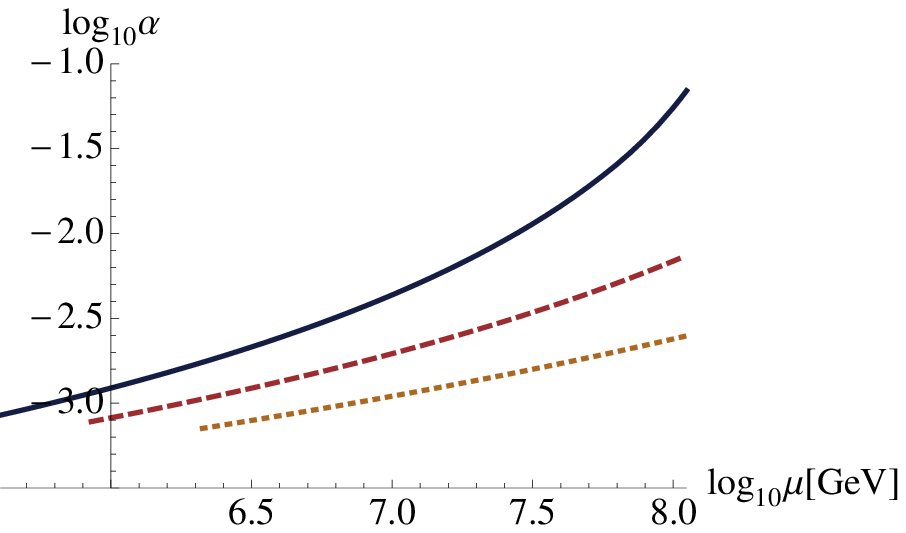}
  \caption{$\alpha$ as a function of $T_*$. 
  The Higgs and top masses are taken to be the same as in Fig.~\ref{fig_lHrun_SM}.}
  \label{fig_alpha_SM}
  \centering
  \includegraphics[width=\columnwidth]{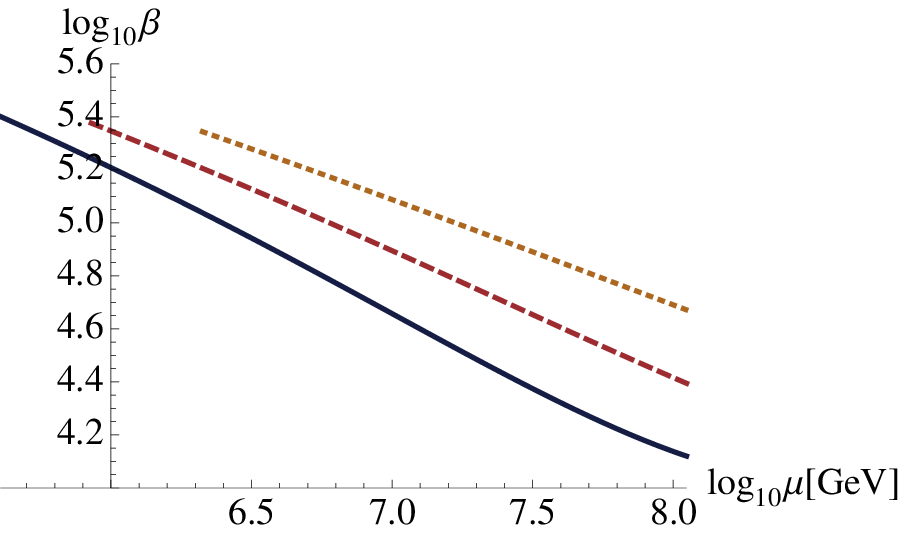}
  \caption{$\beta/H_*$ as a function of $T_*$.
  The Higgs and top masses are taken to be the same as in Fig.~\ref{fig_lHrun_SM}.}
  \label{fig_beta_SM}
\end{figure}

\begin{figure}
  \centering
  \includegraphics[width=\columnwidth]{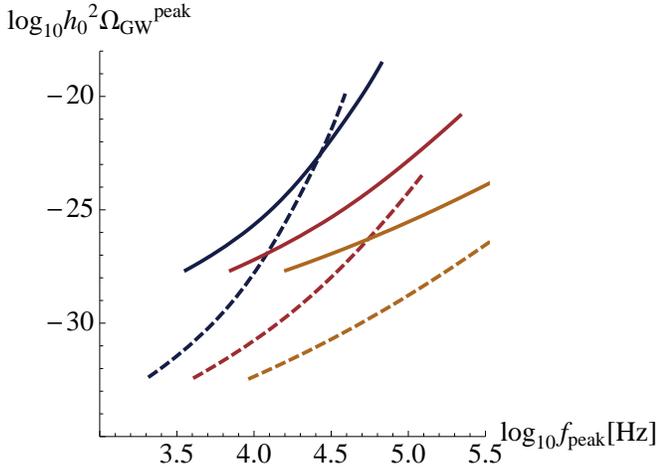}
  \caption{The peak position and amplitude of the GW spectrum for bubble collision (solid lines)
  and turbulence (dotted lines). 
  The Higgs and top masses are taken to be the same as in Fig.~\ref{fig_lHrun_SM}.}
  \label{fig_h2Omega_SM}
\end{figure}

Figure~\ref{fig_h2Omega_SM} shows the peak position and amplitude of the GWs. 
The peak frequency is too high compared to the observable frequency $\sim 1$~Hz;
therefore we need the low frequency behavior to discuss observational possibilities. 
The exponent $n_f$ in the expression
\begin{align}
\Omega_{\rm GW}(f<f_{\rm peak})
&\simeq \Omega_{\rm GW}(f_{\rm peak})\left(\frac{f}{f_{\rm peak}}\right)^{n_{f}}
\end{align}
is roughly $n_{f,{\rm coll}} \simeq 2.6-2.8$ for bubble collision (see, e.g., \cite{Huber:2008hg}), 
though the behavior slightly differs in the literature.
On the other hand, the exponent for turbulence is $n_{f,{\rm turb}} \simeq 2-3$~\cite{Nicolis:2003tg,Megevand:2008mg}.
In any case, it is hard to detect the generated GWs even by the designed future experiments,
e.g., $\Omega_{\rm GW}\sim 10^{-18}$ at the frequency $\sim 1\,$Hz for ultimate DECIGO~\cite{Seto:2001qf}.

There are mainly three reasons for such a smallness of $\Omega_{\rm GW}$ at $f\sim \mathcal{O}(1)$\,Hz.
First, the parameter $\beta/H_\ast$ becomes $\mathcal{O}(10^5)$ in this case. Larger $\beta/H_*$ makes the peak frequency higher and the energy 
fraction $\Omega_{\rm GW}$ lower. Second, there exists a lower limit on the phase transition temperature 
$T_*\gtrsim 10^6$ GeV 
for the first phase transition to take place as mentioned above, which also tends to make the peak frequency high.
The last reason is the smallness of the parameter $\alpha$.

If the Higgs sector is extended, the situation is changed and the detection possibility of GWs may be enhanced.
In the next subsection, we consider the singlet extended Higgs sector as an example of such a new physics.

\subsection{Singlet extension}
\label{se}

In this section, we consider the singlet extended Higgs sector.
In general, if the Higgs field couples to light scalar fields, the generated GWs become stronger
due to the thermal effects.
We consider the two situations depending on the vacuum mass of $S_i$: $m_S^0\simeq \lambda_{S\phi}v_{\rm NP}$. 
The first case is $m_S^{0}\sim T^{\rm PT}_{H}$. 
In this case, the quartic self coupling of the Higgs field $\lambda_H(T)$
is not affected by the singlet sector and we can use the standard model value of $\lambda_H(T)$
around the transition temperature.
The second case is $m^{0}_S\gg T^{\rm PT}_H$, 
where singlets contribute to the running of the Higgs quartic coupling $\lambda_H(T)$, 
and as a result $\lambda_H(T)$ becomes smaller at the transition. 
In such a case the generated GWs can be significantly enhanced as we show later.

\subsubsection{The case with $m_S^{0}\sim T^{\rm PT}_{H}$}

\begin{figure}
  \centering
  \includegraphics[width=0.9\columnwidth]{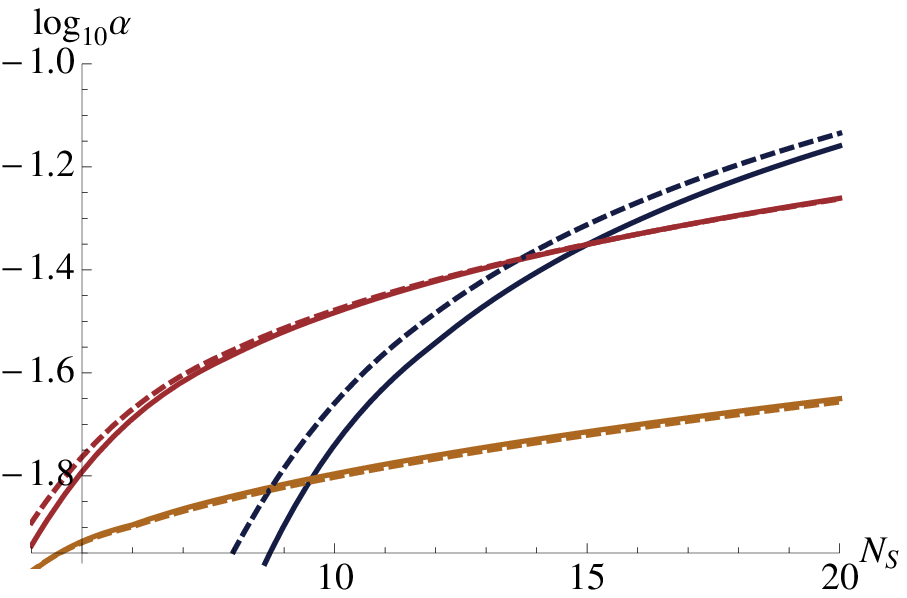}
  \caption{$\alpha$ with $f_{\rm peak}=1$\,Hz as a function of $N_S$.
  Solid lines correspond to bubble collision, while dotted lines correspond to turbulence.
  $\lambda_{SH}=1$ (blue), $1.5$ (red) and $2$ (yellow).}
  \label{fig_alpha_singlet_1}
  \centering
  \includegraphics[width=0.9\columnwidth]{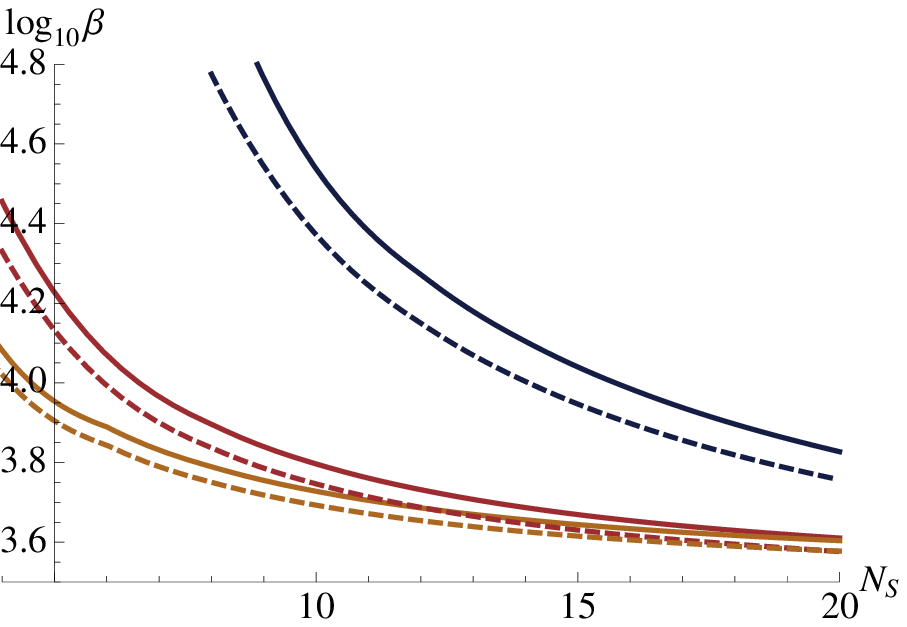}
  \caption{
  $\beta/H_*$ with $f_{\rm peak}=1$\,Hz as a function of $N_S$.
  Solid lines correspond to bubble collision, while dotted lines correspond to turbulence.
  $\lambda_{SH}=1$ (blue), $1.5$ (red) and $2$ (yellow).
  }
  \label{fig_beta_singlet_1}
\end{figure}

\begin{figure}
  \centering
  \includegraphics[width=0.9\columnwidth]{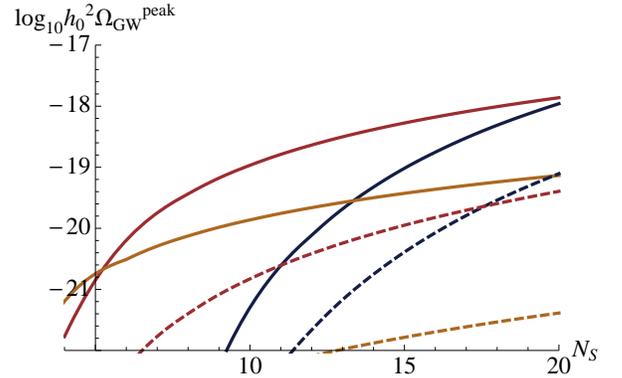}
  \caption{
  The energy fraction $\Omega_{\rm GW}$ with $f_{\rm peak}=1$\,Hz as a function of $N_S$.
  Each line corresponds to 
  $\lambda_{SH}=1$ (solid blue), $1.5$ (solid red), and $2$ (solid yellow) for bubble collision 
  (\ref{eq_h2Omega_coll}),
  and 
  $\lambda_{SH}=1$ (blue dashed), $1.5$ (red dashed), and $2$ (yellow dashed) for turbulence 
  (\ref{eq_h2Omega_tur}).
  }
  \label{fig_h2Omega_singlet_1}
\end{figure}

With additional singlets, 
the first order phase transition of the Higgs field 
can occur 
even below $\sim 10^6\,$GeV
if $\lambda_H/\lambda_{SH}^2$ is small enough.
In order to show the typical strength of the GWs, we fix the peak frequency at $f_{\rm peak}=1$\,Hz.
Figures~\ref{fig_alpha_singlet_1}--\ref{fig_beta_singlet_1} show
$\alpha$ and $\beta/H_*$ as a function of the number of singlets, respectively.
Also, Fig.~\ref{fig_h2Omega_singlet_1} shows the energy fraction $\Omega_{\rm GW}$ with $f_{\rm peak}=1$\,Hz. 
The blue, red and yellow lines correspond to the case with 
$\lambda_{SH}=1, 1.5, 2$, respectively.\footnote{
As long as $N \lambda_{SH}^4/16\pi^2 \lesssim {\mathcal O}(1)$ and there are no interactions among $S_i$s,
the higher order corrections on the potential are not important.
}
It is seen that for large enough $N_S\gtrsim 20$, $\Omega_{\rm GW}$ can become $\sim 10^{-18}$,
which may be within the sensitivity of future experiments~\cite{Seto:2001qf}.
\footnote{
The calculation of the strength of the GWs
in singlet extensions is done in~\cite{Espinosa:2008kw,Kehayias:2009tn,Leitao:2012tx,Kakizaki:2015wua}.
In these studies, the strength of the GWs is more enhanced for the large $N_S$ and $\lambda_{SH}$ region.
The difference comes from the treatment of the zero temperature potential.
}

\subsubsection{The case with $m_S^{0}\gg T^{\rm PT}_{H}$}

\begin{figure}
  \centering
  \includegraphics[width=\columnwidth]{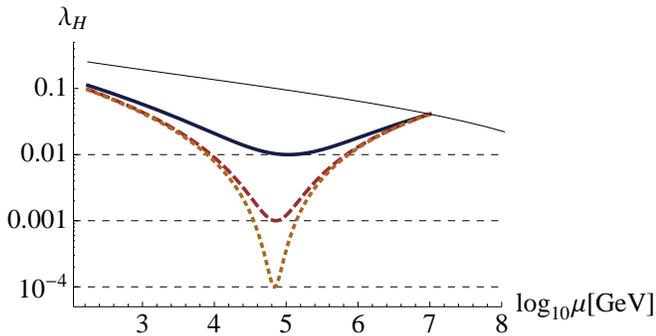}
  \caption{
  Running of the Higgs quartic coupling $\lambda_H$.
  Parameters are taken to be $m_S = 10^7$~GeV, and 
  $\lambda_{H,{\rm min}} = 10^{-2}$ (solid blue), $10^{-3}$ (red dashed), $10^{-4}$ (yellow dotted). 
  The black line corresponds to the running without singlet.
  }
  \label{fig_lHrun_singlet_2}
\end{figure}

Now, let us consider the case with $m_{S}^{0}\gg T^{\rm PT}_H$.
In this case, the self-quartic coupling of the Higgs field $\lambda_H$ is different from the standard model value at $T^{\rm PT}_H$.
At zero temperature, the running of the couplings is the same as the standard model one for $\mu<m_S^0$ with
$\mu$ being the renormalization scale.
On the other hand, when $\phi_{\rm NP}$ is trapped at the origin at high temperature,
the coupling $\lambda_{SH}$ affects the running of the couplings, especially $\lambda_H$. 
At the one-loop level, the renormalization group equations become (see, e.g., \cite{Costa:2014qga})
\begin{align}
	\frac{d\lambda_H}{d\ln \mu}&=\beta_H^{\rm SM}+\frac{N_S}{16\pi^2}\lambda_{SH}^4,\\
	\frac{d\lambda_{SH}}{d\ln \mu}&=\frac{\lambda_{SH}}{16\pi^2}\left[
	2\lambda_{SH}^2+3y_t^2-\frac{3}{4}g'^2-\frac{9}{4}g_2^2
	\right],
\end{align}
where $\beta^{\rm SM}$ denotes the standard model contribution.
Figure~\ref{fig_lHrun_singlet_2} shows the running of $\lambda_H$ with $\lambda_{SH}$ (black) 
and without $\lambda_{SH}$ (blue, red, and yellow).
We set $m_S^0=10^7$ GeV and $N_S=4$, 
and $\lambda_{SH}(m_S^0) \simeq 1$ is chosen so that the minimal value of $\lambda_H$
becomes $10^{-2},10^{-3},10^{-4}$ in blue, red, and yellow lines, respectively.
Note that the strength of the produced GWs becomes stronger for smaller $\lambda_H$.

\begin{figure}
  \centering
  \includegraphics[width=\columnwidth]{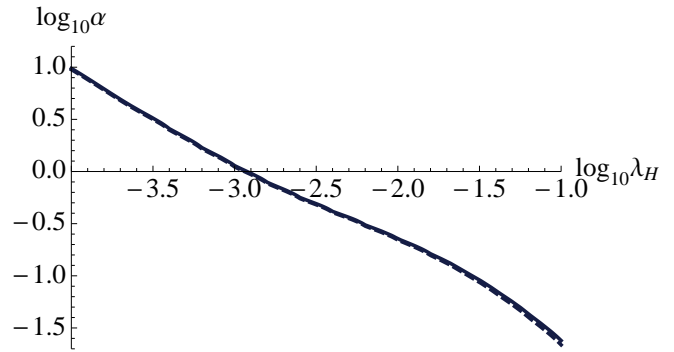}
  \caption{
  $\alpha$ as a function of $\lambda_H$.
  Each line corresponds to bubble collision (solid) and turbulence (dashed).
  }
  \label{fig_alpha_singlet_2}
  \end{figure}

\begin{figure}
  \centering
  \includegraphics[width=\columnwidth]{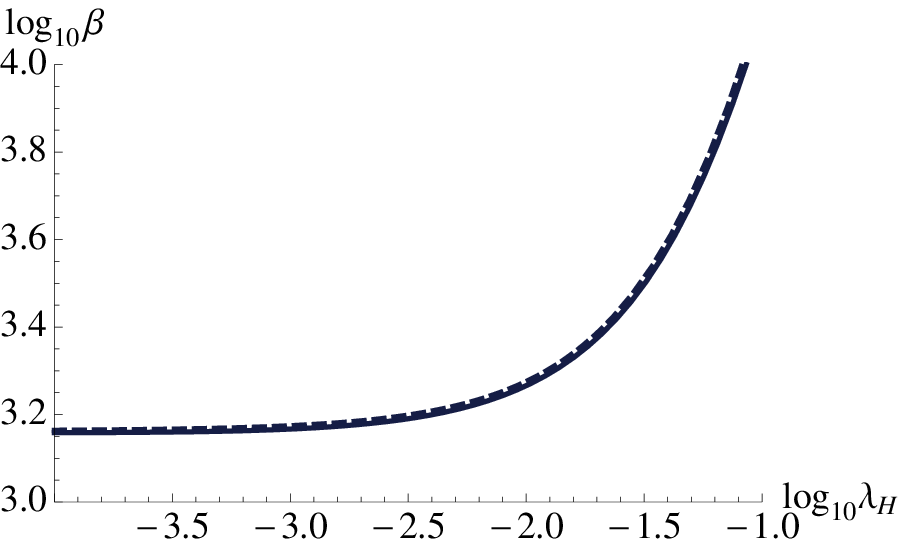}
  \caption{
  $\beta/H_*$ as a function of $\lambda_H$.
  Each line corresponds to bubble collision (solid) and turbulence (dashed).
  }
  \label{fig_beta_singlet_2}
\end{figure}

\begin{figure}
  \centering
  \includegraphics[width=\columnwidth]{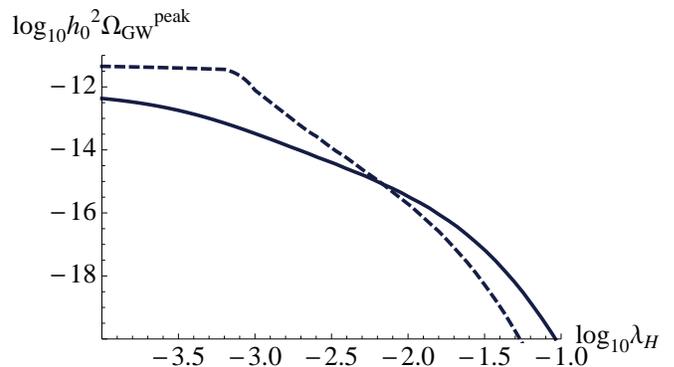}
  \caption{
  GW energy fraction $\Omega_{\rm GW}$ as a function of $\lambda_*$ 
  in the case of $m_S^{0}\gg T^{\rm PT}_{H}$. 
  Each line corresponds to bubble collision (solid) and turbulence (dashed).
  See the text for details.
  }
  \label{fig_h2Omega_singlet_2}
\end{figure}

\begin{figure}
  \centering
  \includegraphics[width=\columnwidth]{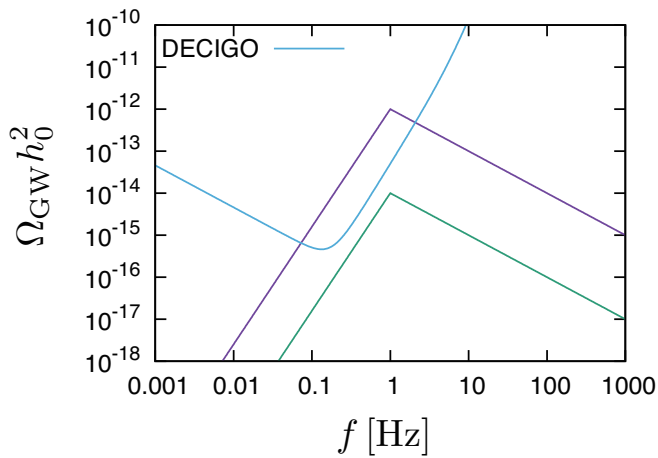}
  \caption{The GW spectrum from bubble collisions for $\Omega_{\rm GW}(f_{\rm peak})=10^{-12}$ and $10^{-14}$
  with $f_{\rm peak}=1$\,Hz. Together shown is the sensitivity of the DECIGO.
  }
  \label{fig:spec}
\end{figure}

In order to see the typical situation where the GW amplitude is significantly enhanced,
we assume that the phase transition occurs at the point where 
$\lambda_H$ takes its minimal value $\lambda_{H,{\rm min}}$, i.e. at
\begin{align}
	\frac{d\lambda_H}{d\ln \mu}&=\beta_H^{\rm SM}+\frac{N_S}{16\pi^2}\lambda_{SH}^4=0.
\end{align}
Figures~\ref{fig_alpha_singlet_2}--\ref{fig_h2Omega_singlet_2} 
show the parameter $\alpha$, $\beta/H_*$ and the GW energy fraction $\Omega_{\rm GW}$ 
at the frequency 1\,Hz, respectively.\footnote{
Because of the smallness of $\lambda_H$, the field value of the Higgs field after the transition becomes relatively large.
In such a situation, the density of $S_i$ particles is supposed to be suppressed.
This may cause subsequent transition of $\phi_{\rm NP}$ field because the thermal mass of $\phi_{\rm NP}$ becomes small.
In such a situation, the strength of the GWs gets enhanced because
the parameter $\alpha$ becomes larger.
}
We have taken $N_S=4$ and $f_{\rm peak}=1$\,Hz.
For $\lambda_{H,{\rm min}}\lesssim 0.01$, $\Omega_{\rm GW}$ can be greater than $\sim 10^{-15}$.
This is within a sensitivity of future experiments~\cite{Seto:2001qf,Kuroyanagi:2014qza}.
Figure~\ref{fig:spec} shows the GW spectrum from bubble collisions for $\Omega_{\rm GW}(f_{\rm peak})=10^{-12}$ and $10^{-14}$
with $f_{\rm peak}=1$\,Hz. Together shown is the sensitivity of the DECIGO.
Note that there are huge foreground GWs from white dwarf binaries below $\sim 0.1-1$\,Hz~\cite{Farmer:2003pa},
but still GWs  from the phase transition may be observable.

\section{Conclusions}
\label{Dis}

In this paper, we have considered GWs generated by the first order phase transition of the Higgs field 
at some new physics scale.
If the new physics contains scalar fields $(\phi_{\rm NP})$, 
the couplings between the standard model Higgs field and such scalars exist in general.
These couplings can cause the first order phase transition of the Higgs field 
at the temperature of the Universe around the new physics scale, which is much higher than the weak scale.
Hence the peak position of the GWs as well as their strength can take a broad range of values 
depending on the new physics scale. 

We considered two types of models in the Higgs sector.
In the first model we have only the standard model Higgs boson and $\phi_{\rm NP}$.
In this case we have seen that the generated GW is too weak to detect by designed future experiments.
The second model contains additional singlet fields $S_i$ 
and we have shown that the detection of the GWs may be possible if
the number of the singlets is $\mathcal{O}(10)$ or the self-quartic coupling of the Higgs field 
$\lambda_H$ is small enough $\lesssim 0.01$
due to the coupling of the Higgs boson with additional singlets.

As a final remark, in this paper we considered GWs associated with first order phase transition 
of the standard model Higgs field,
which happens at a much higher scale than the weak scale.
There are also possibilities that the phase transition of some other scalar fields that do or do not couple 
to the Higgs field is first order and generates GWs strong enough to be detected.
Since the scale of phase transition of these fields need not be the electroweak scale, it may open up a new possibility 
for probing new physics through GW detection.

\section*{ACKNOWLEDGEMENTS}

This work was supported by Grant-in-Aid for Scientific Research Grants No.~26104009 (K.N.), 
No.~15H05888 (K.N.), No.~26247042 (K.N.), No.~26800121 (K.N.), MEXT, Japan.
The work of R.J. and M.T. is supported in part by JSPS Research Fellowships
for Young Scientists and by the Program for Leading Graduate Schools, MEXT, Japan.




\begin{thebibliography}{99}

\bibitem{Starobinsky:1979ty} 
  A.~A.~Starobinsky,
  JETP Lett.\  {\bf 30}, 682 (1979)
  [Pisma Zh.\ Eksp.\ Teor.\ Fiz.\  {\bf 30}, 719 (1979)].

\bibitem{Vilenkin:2000jqa} 
  A.~Vilenkin and E.~P.~S.~Shellard,
  ``{\it Cosmic Strings and Other Topological Defects},"
Cambridge Monographs on Mathematical Physics 
(Cambridge University Press, Cambridge, 2000), 
ISBN:9780521654760.
 
\bibitem{Witten:1984rs} 
  E.~Witten,
  Phys.\ Rev.\ D {\bf 30}, 272 (1984).
  
\bibitem{Hogan:1986qda} 
  C.~J.~Hogan,
  Mon.\ Not.\ Roy.\ Astron.\ Soc.\  {\bf 218}, 629 (1986).
 
\bibitem{Turner:1990rc} 
  M.~S.~Turner and F.~Wilczek,
  Phys.\ Rev.\ Lett.\  {\bf 65}, 3080 (1990).
  
\bibitem{Kosowsky:1991ua} 
  A.~Kosowsky, M.~S.~Turner and R.~Watkins,
  Phys.\ Rev.\ D {\bf 45}, 4514 (1992).
  
\bibitem{Kosowsky:1992rz} 
  A.~Kosowsky, M.~S.~Turner and R.~Watkins,
  Phys.\ Rev.\ Lett.\  {\bf 69}, 2026 (1992).

\bibitem{Turner:1992tz} 
  M.~S.~Turner, E.~J.~Weinberg and L.~M.~Widrow,
  Phys.\ Rev.\ D {\bf 46}, 2384 (1992).
  
\bibitem{Kosowsky:1992vn}
  A.~Kosowsky and M.~S.~Turner,
  Phys.\ Rev.\ D {\bf 47} (1993) 4372
  [astro-ph/9211004].

\bibitem{Kamionkowski:1993fg} 
  M.~Kamionkowski, A.~Kosowsky and M.~S.~Turner,
  Phys.\ Rev.\ D {\bf 49}, 2837 (1994)
  [astro-ph/9310044].
 
\bibitem{Seoane:2013qna} 
  P.~A.~Seoane {\it et al.} [eLISA Collaboration],
  arXiv:1305.5720 [astro-ph.CO].
  
\bibitem{Harry:2006fi} 
  G.~M.~Harry, P.~Fritschel, D.~A.~Shaddock, W.~Folkner and E.~S.~Phinney,
Class.\ Quant.\ Grav.\  {\bf 23}, 4887 (2006)
[Erratum-ibid.\  {\bf 23}, 7361 (2006)].

\bibitem{Seto:2001qf} 
  N.~Seto, S.~Kawamura and T.~Nakamura,
  Phys.\ Rev.\ Lett.\  {\bf 87}, 221103 (2001)
  [astro-ph/0108011].

\bibitem{Harry:2010zz} 
  G.~M.~Harry [LIGO Scientific Collaboration],
  Class.\ Quant.\ Grav.\  {\bf 27}, 084006 (2010).

\bibitem{Somiya:2011np} 
  K.~Somiya [KAGRA Collaboration],
  Class.\ Quant.\ Grav.\  {\bf 29}, 124007 (2012)
  [arXiv:1111.7185 [gr-qc]].
  
\bibitem{TheVirgo:2014hva} 
  F.~Acernese {\it et al.} [VIRGO Collaboration],
  Class.\ Quant.\ Grav.\  {\bf 32}, no. 2, 024001 (2015)
  [arXiv:1408.3978 [gr-qc]].
  
\bibitem{Dine:1992wr} 
  M.~Dine, R.~G.~Leigh, P.~Y.~Huet, A.~D.~Linde and D.~A.~Linde,
  Phys.\ Rev.\ D {\bf 46}, 550 (1992)
  [hep-ph/9203203].
  
\bibitem{Farakos:1994kx} 
  K.~Farakos, K.~Kajantie, K.~Rummukainen and M.~E.~Shaposhnikov,
  Nucl.\ Phys.\ B {\bf 425}, 67 (1994)
  [hep-ph/9404201].
  
\bibitem{Arnold:1992rz} 
  P.~B.~Arnold and O.~Espinosa,
  Phys.\ Rev.\ D {\bf 47}, 3546 (1993)
  [Phys.\ Rev.\ D {\bf 50}, 6662 (1994)]
  [hep-ph/9212235].
  
\bibitem{Fodor:1994bs} 
  Z.~Fodor and A.~Hebecker,
  Nucl.\ Phys.\ B {\bf 432}, 127 (1994)
  [hep-ph/9403219].
  
\bibitem{Kajantie:1995kf} 
  K.~Kajantie, M.~Laine, K.~Rummukainen and M.~E.~Shaposhnikov,
  Nucl.\ Phys.\ B {\bf 466}, 189 (1996)
  [hep-lat/9510020].
  
\bibitem{Karsch:1996aw} 
  F.~Karsch, T.~Neuhaus, A.~Patkos and J.~Rank,
  Nucl.\ Phys.\ B {\bf 474}, 217 (1996)
  [hep-lat/9603004].
  
\bibitem{Kajantie:1996mn} 
  K.~Kajantie, M.~Laine, K.~Rummukainen and M.~E.~Shaposhnikov,
  Phys.\ Rev.\ Lett.\  {\bf 77}, 2887 (1996)
  [hep-ph/9605288].
  
\bibitem{Kajantie:1996qd} 
  K.~Kajantie, M.~Laine, K.~Rummukainen and M.~E.~Shaposhnikov,
  Nucl.\ Phys.\ B {\bf 493}, 413 (1997)
  [hep-lat/9612006].
    
\bibitem{Gurtler:1997hr} 
  M.~Gurtler, E.~M.~Ilgenfritz and A.~Schiller,
  Phys.\ Rev.\ D {\bf 56}, 3888 (1997)
  [hep-lat/9704013].
  
\bibitem{Rummukainen:1998as} 
  K.~Rummukainen, M.~Tsypin, K.~Kajantie, M.~Laine and M.~E.~Shaposhnikov,
  Nucl.\ Phys.\ B {\bf 532}, 283 (1998)
  [hep-lat/9805013].
  
\bibitem{Csikor:1998eu} 
  F.~Csikor, Z.~Fodor and J.~Heitger,
  Phys.\ Rev.\ Lett.\  {\bf 82}, 21 (1999)
  [hep-ph/9809291].
  
\bibitem{Aoki:1999fi} 
  Y.~Aoki, F.~Csikor, Z.~Fodor and A.~Ukawa,
  Phys.\ Rev.\ D {\bf 60}, 013001 (1999)
  [hep-lat/9901021].
  
\bibitem{Peccei:1977hh} 
  R.~D.~Peccei and H.~R.~Quinn,
  Phys.\ Rev.\ Lett.\  {\bf 38}, 1440 (1977).

\bibitem{Kim:1986ax} 
  J.~E.~Kim,
  Phys.\ Rept.\  {\bf 150}, 1 (1987).
  
\bibitem{Moroi:2013tea} 
  T.~Moroi, K.~Mukaida, K.~Nakayama and M.~Takimoto,
  JHEP {\bf 1306}, 040 (2013)
  [arXiv:1304.6597 [hep-ph]];
  JHEP {\bf 1411}, 151 (2014)
  [arXiv:1407.7465 [hep-ph]].
  
\bibitem{Hindmarsh:2013xza} 
  M.~Hindmarsh, S.~J.~Huber, K.~Rummukainen and D.~J.~Weir,
  Phys.\ Rev.\ Lett.\  {\bf 112}, 041301 (2014)
  [arXiv:1304.2433 [hep-ph]].
  
\bibitem{Hindmarsh:2015qta} 
  M.~Hindmarsh, S.~J.~Huber, K.~Rummukainen and D.~J.~Weir,
  arXiv:1504.03291 [astro-ph.CO].
  
\bibitem{Kalaydzhyan:2014wca} 
  T.~Kalaydzhyan and E.~Shuryak,
  Phys.\ Rev.\ D {\bf 91}, no. 8, 083502 (2015)
  [arXiv:1412.5147 [hep-ph]].
  
\bibitem{Caprini:2007xq} 
  C.~Caprini, R.~Durrer and G.~Servant,
  Phys.\ Rev.\ D {\bf 77}, 124015 (2008)
  [arXiv:0711.2593 [astro-ph]].
  
\bibitem{Caprini:2009yp} 
  C.~Caprini, R.~Durrer and G.~Servant,
  JCAP {\bf 0912}, 024 (2009)
  [arXiv:0909.0622 [astro-ph.CO]].
  
\bibitem{Caprini:2010xv} 
  C.~Caprini, R.~Durrer and X.~Siemens,
  Phys.\ Rev.\ D {\bf 82}, 063511 (2010)
  [arXiv:1007.1218 [astro-ph.CO]].

\bibitem{Nicolis:2003tg}
  A.~Nicolis,
  Class.\ Quant.\ Grav.\  {\bf 21}, L27 (2004)
  [gr-qc/0303084].

\bibitem{Caprini:2006jb} 
  C.~Caprini and R.~Durrer,
  Phys.\ Rev.\ D {\bf 74}, 063521 (2006)
  [astro-ph/0603476].

\bibitem{Huber:2008hg} 
  S.~J.~Huber and T.~Konstandin,
  JCAP {\bf 0809}, 022 (2008)
  [arXiv:0806.1828 [hep-ph]].

\bibitem{Giblin:2014qia} 
  J.~T.~Giblin and J.~B.~Mertens,
  Phys.\ Rev.\ D {\bf 90}, no. 2, 023532 (2014)
  [arXiv:1405.4005 [astro-ph.CO]].

\bibitem{Dolgov:2002ra} 
  A.~D.~Dolgov, D.~Grasso and A.~Nicolis,
  Phys.\ Rev.\ D {\bf 66}, 103505 (2002)
  [astro-ph/0206461].
  
\bibitem{Gogoberidze:2007an} 
  G.~Gogoberidze, T.~Kahniashvili and A.~Kosowsky,
  Phys.\ Rev.\ D {\bf 76}, 083002 (2007)
  [arXiv:0705.1733 [astro-ph]].

\bibitem{Steinhardt:1981ct} 
  P.~J.~Steinhardt,
  Phys.\ Rev.\ D {\bf 25}, 2074 (1982).

\bibitem{Bodeker:2009qy} 
  D.~Bodeker and G.~D.~Moore,
  JCAP {\bf 0905}, 009 (2009)
  [arXiv:0903.4099 [hep-ph]].
  
\bibitem{Kozaczuk:2015owa} 
  J.~Kozaczuk,
  arXiv:1506.04741 [hep-ph].

\bibitem{Coleman:1977py} 
  S.~R.~Coleman,
  Phys.\ Rev.\ D {\bf 15}, 2929 (1977)
  [Phys.\ Rev.\ D {\bf 16}, 1248 (1977)].
  
\bibitem{Callan:1977pt} 
  C.~G.~Callan, Jr. and S.~R.~Coleman,
  Phys.\ Rev.\ D {\bf 16}, 1762 (1977).
  
\bibitem{Linde:1981zj} 
  A.~D.~Linde,
  Nucl.\ Phys.\ B {\bf 216}, 421 (1983)
  [Nucl.\ Phys.\ B {\bf 223}, 544 (1983)].
  
\bibitem{Quiros:1994dr} 
  M.~Quiros,
  Helv.\ Phys.\ Acta {\bf 67}, 451 (1994).
  
\bibitem{Agashe:2014kda} 
  K.~A.~Olive {\it et al.} [Particle Data Group Collaboration],
  Chin.\ Phys.\ C {\bf 38}, 090001 (2014).
  
\bibitem{Aad:2015zhl} 
  G.~Aad {\it et al.} [ATLAS and CMS Collaborations],
  Phys.\ Rev.\ Lett.\  {\bf 114}, 191803 (2015)
  [arXiv:1503.07589 [hep-ex]].
  
\bibitem{ATLAS:2014wva} 
  [ATLAS and CDF and CMS and D0 Collaborations],
  arXiv:1403.4427 [hep-ex].
  
\bibitem{Megevand:2008mg} 
  A.~Megevand,
  Phys.\ Rev.\ D {\bf 78}, 084003 (2008)
  [arXiv:0804.0391 [astro-ph]].

  
  
  
\bibitem{Espinosa:2008kw} 
  J.~R.~Espinosa, T.~Konstandin, J.~M.~No and M.~Quiros,
  Phys.\ Rev.\ D {\bf 78}, 123528 (2008)
  [arXiv:0809.3215 [hep-ph]].
  
\bibitem{Kehayias:2009tn} 
  J.~Kehayias and S.~Profumo,
  JCAP {\bf 1003}, 003 (2010)
  [arXiv:0911.0687 [hep-ph]].
  
\bibitem{Leitao:2012tx} 
  L.~Leitao, A.~Megevand and A.~D.~Sanchez,
  JCAP {\bf 1210}, 024 (2012)
  [arXiv:1205.3070 [astro-ph.CO]].
  
\bibitem{Kakizaki:2015wua} 
  M.~Kakizaki, S.~Kanemura and T.~Matsui,
  arXiv:1509.08394 [hep-ph].
  
  
\bibitem{Costa:2014qga} 
  R.~Costa, A.~P.~Morais, M.~O.~P.~Sampaio and R.~Santos,
  Phys.\ Rev.\ D {\bf 92}, no. 2, 025024 (2015)
  [arXiv:1411.4048 [hep-ph]].
  
\bibitem{Kuroyanagi:2014qza} 
  S.~Kuroyanagi, K.~Nakayama and J.~Yokoyama,
  PTEP {\bf 2015}, no. 1, 013E02 (2015)
  [arXiv:1410.6618 [astro-ph.CO]].
  
\bibitem{Farmer:2003pa} 
  A.~J.~Farmer and E.~S.~Phinney,
  Mon.\ Not.\ Roy.\ Astron.\ Soc.\  {\bf 346}, 1197 (2003)
  [astro-ph/0304393].


\end{thebibliography}
\end{document}